# Towards Agentic Test-Driven Quality Assurance for 6G Networks

Christos Tranoris, Besiana Agko
Electrical and Computing Eng. Dept,
University of Patras, Patras, Greece
tranoris@ece.upatras.gr, besiana_agko@ac.upatras.gr

Kostis Trantzas, Irene Denazi
NetonomIQ PC
Patras, Greece
ktrantzas@netonomiq.com, idenazi@netonomiq.com

*Abstract*—This work proposes an agentic, intent-driven end-to-end (E2E) orchestration framework that integrates intent co-creation with a Test-Driven Quality Assurance paradigm. In this framework, autonomous agents iteratively refine a user's initial intent into a confirmed, auditable specification. Furthermore, the system automatically derives validation tests from these intents before provisioning, directly mirroring the Test-Driven Development workflow in software engineering to ensure proactive Service Level Agreement (SLA) compliance. The architecture is grounded in a standards-aligned knowledge representation using TM Forum (TMF) information models and catalogs. This enables deterministic graph traversal from high-level Product Offerings down to granular Service/Resource and Test specifications. We prototyped this architecture by extending OpenSlice with a message-driven, multi-agent pattern and integrating MCP-enabled (Model Context Protocol) tool access for real-time knowledge retrieval. Currently, our evaluation of the agents targets the intent co-creation phase as a baseline toward full-scale orchestration. Building on experiments with multiple open-source Large Language Model (LLM) backends integrated with the TMF-based knowledge base, we observe substantial variability in tool-use reliability and hallucination patterns, underscoring the critical importance of robust knowledge integration in agentic 6G systems.

## I. Introduction

6G networks will be far more heterogeneous across radio, transport, core, and cloud–edge domains, making manual or semi-automated operations inadequate for dynamic provisioning, ultra-low latency services, and stringent reliability/sustainability targets. As a result, networks must progress toward high autonomy (Levels 4–5) per TM Forum/3GPP [1], [2]. Intent-Based Networking (IBN) facilitates this transition by expressing what outcomes are desired rather than how to configure the network [3]. While contemporary IBN assumes static intents that map directly to pre-defined workflows, real-world requirements are inherently ambiguous or incomplete. Furthermore, traditional assurance is typically reactive, executed only after provisioning, which delays the detection of configuration mismatches and Service Level Agreement (SLA) violations.[4]. Agentic AI addresses these gaps by reasoning over multi-domain data, coordinating across specialized agents, and adapting via iterative feedback loops. However, this autonomy introduces new challenges regarding trust, governance, and standards alignment [5]. To tackle these challenges, we propose an agentic, intent-driven end-to-end (E2E) orchestration framework that (i) enables intent co-creation, where autonomous agents iteratively enrich and refine ambiguous user inputs with operational, quality, and cost constraints to obtain an explicit, auditable confirmation; and (ii) applies a test-driven quality assurance paradigm. This model automatically derives Service Level Objective (SLO)/SLA validation tests from the confirmed intent before provisioning, ensuring that deployment "earns" compliance progressively. This approach is analogous to Test-Driven Development (TDD) in Software Engineering, where validation tests initially fail and are only satisfied as the system is successfully implemented and configured via the E2E service orchestrator.

## II. Related Work

The transition to Future 6G networks necessitates orchestrating highly heterogeneous services with strict and volatile requirements (latency, reliability, throughput, cost, sustainability) across multiple administrative and technology domains, while still delivering predictable and verifiable E2E quality [7]. Despite the push for AI-native automation, three critical gaps remain in the current state of the art: (i) contemporary IBN often assumes "well-formed" intents that map linearly to workflows. However, real user/operator intents are often incomplete or ambiguous, especially for new services or runtime modifications, so the former assumption leads to brittle mappings, manual fixes, or unmet expectations [7], [8]; (ii) a semantic mismatch between human-centric intent descriptions and machine-executable actions. Template or rule-based translators struggle to adapt to new service types, cross-domain trade-offs, and evolving conditions [9]; and (iii) Quality Assurance (QA) is traditionally treated as a separate, post-deployment activity. This "provision-then-verify" model expects SLAs/SLOs and validation to be defined offline, rendering verification reactive and preventing quality objectives from guiding provisioning decisions upfront [10].

At the same time, while AI and agent-based methods offer a path to managing complexity, their integration into operational orchestration remains fraught with tension. Direct AI control raises concerns about determinism, trust, and standards compliance, whereas restricting AI to passive analytics underutilizes its potential for reasoning, negotiation, and adaptation in E2E scenarios [11]. This complexity motivates the central research problem, i.e., designing an orchestration framework capable of interactively refining intents, translating them into deterministic executable actions, and continuously verifying quality, while maintaining compatibility with legacy control systems and established standards. Addressing this challenge requires a unified approach that (i) facilitates intent co-creation and validation to resolve ambiguity, (ii) bridges the semantic gap between high-level intent and low-level action, and (iii) elevates quality specification and test-driven validation to first-class citizens throughout the service lifecycle. Ultimately, such a framework must leverage agentic AI for cross-domain coordination without compromising the governability or trust essential for carrier-grade 6G operations.

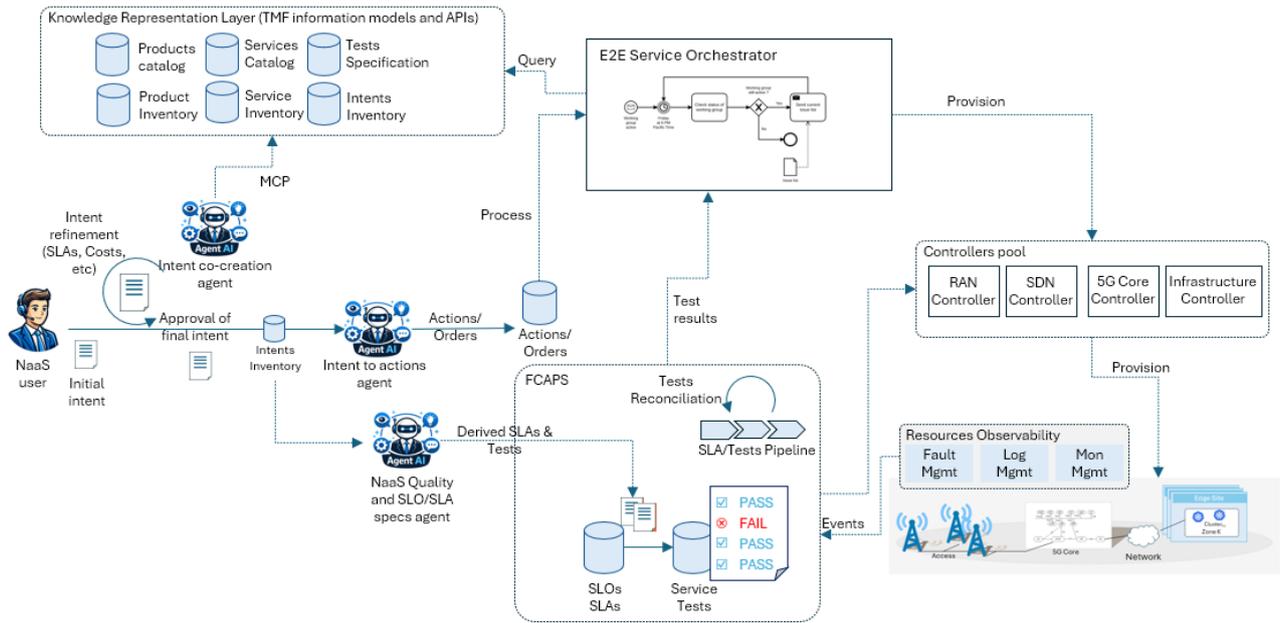

*Figure 1 Proposed Agentic, Intent-Driven E2E Orchestration for Quality Assurance*

The semantic gap between high-level user intent and machine-executable actions can be resolved by utilizing a standards-aligned knowledge representation. Communication Service Providers (CSPs) commercialize a network service as a TM Forum Product Offering (TMF620). This structured catalog entity allows for customer orders, bundling cost, policies, and SLA terms, and links to one or more Product Specifications, which serve as the formal template for technical characteristics and configurable parameters (e.g., location, capacity, duration) [12]. When an order is placed, the orchestration stack decomposes it into granular Service and Resource orders, coordinating workflows via an E2E orchestrator and specialized domain controllers. To ensure operational continuity, the framework maintains runtime traceability through Product, Service, and Resource Inventories (TMF637, TMF638, TMF639). This standards-aligned structure enables full lifecycle operations, including activation, scaling, modification, healing, and termination.

In this framework, assurance is seamlessly integrated via standardized test management (TMF653), which enables the definition of Service Test Specifications [13]. These specifications allow for the execution of diverse probes, including connectivity, QoS (latency/throughput), slice admission, and API availability, both as pre-activation acceptance checks and as continuous runtime validation. The resulting auditable Service Test Results are explicitly linked to service instances [13], providing the evidentiary basis for compliance. Consequently, SLO/SLA management ties business commitments to these measurable outcomes, synthesizing real-time telemetry with TMF653 results to verify compliance and trigger closed-loop remediation if targets drift or tests fail. This TM Forum-aligned chain, spanning from Product Offering and Specifications through Orchestration and Inventories to TMF653 Testing and SLO/SLA Compliance, establishes a controlled, standards-based foundation to deliver and continuously manage E2E network services.

*A. Intent-Based Networking and Autonomous Networks*

IBN has been widely studied as a means to simplify network management by abstracting low-level configuration details. Standardization bodies, such as the TM Forum [14] and ETSI [15], as well as contemporary research [16], have positioned intent management frameworks as core components of Autonomous Network (AN) architectures. These frameworks typically employ closed-loop control to translate high-level intents into actionable policies and configurations across heterogeneous network domains. Furthermore, assurance mechanisms remain largely decoupled from intent processing, reducing the system's ability to verify intent fulfillment in a timely and systematic manner.

*B. AI and Agent-Based Network Management*

The convergence of Agentic AI and Multi-Agent Systems (MAS) has fundamentally transformed AN control, with a specific focus on distributed, large-scale systems, such as modern cellular networks [17]. AI techniques have been increasingly applied to network management tasks, including optimization, fault detection, and predictive analytics. More recently, agent-based and multi-agent systems have been proposed to enable distributed intelligence across network domains. In these approaches, autonomous agents observe the network states, reason about objectives, and coordinate actions through inter-agent communication [18] [19].

Agentic AI frameworks evolve this concept by leveraging Large Language Models (LLMs) and advanced reasoning chains. This enables agents to interpret high-level goals, negotiate with peers, and adapt their strategies over time. Research efforts in Open RAN, AI-native networks, and AI-driven orchestration have demonstrated the potential of such systems to improve flexibility and responsiveness. However, most existing works focus on isolated domains (e.g., RAN optimization) or assume direct agent control over network functions. Such unmediated control raises concerns about determinism, operational safety, and compatibility with legacy systems. Critical challenges remain in conflict mitigation, secure decision-making, and the guarantee of trustable and explainable AI [20].

### C. Service Assurance and Test-Driven Approaches

Service assurance traditionally focuses on monitoring and validating network behavior post-deployment, utilizing KPIs, alarms, and SLA reports. In contrast, software engineering has long adopted TDD [21], a practice where automated tests are defined before implementation to guide design and validate correctness throughout development and eventually ensure the delivery of high-quality products. The TDD workflow typically follows a short iterative cycle: (i) specifying desired behavior through a test that initially fails, (ii) implementing the minimal logic required to pass the test, and (iii) refactoring the implementation to improve structure, while maintaining "green" status.

By forcing requirements to be expressed as executable checks upfront, TDD reduces ambiguity, promotes modular and maintainable code, and provides a continuously runnable regression suite that increases confidence in changes over time. To the best of our knowledge, no existing work has explored the application of these TDD principles to the domain of 6G network and service orchestration. Existing approaches typically define tests manually or as part of post-deployment validation pipelines. Current approaches still define tests manually or as part of isolated post-deployment pipelines, leaving quality objectives loosely coupled with user intent and rendering assurance reactive rather than proactive.

### III. OUR APPROACH

#### A. Design Principles

Figure 1 depicts our proposed architecture for Agentic, Intent-Driven E2E Orchestration with the goal of Test-Driven Quality Assurance on delivering Network services. The framework initiates a paradigm shift from manual scripting to a collaborative, AI-driven intake process. Here, a specialized Intent *Co-creation Agent* serves as the primary interface, engaging the user in a refinement loop to align high-level business objectives with feasible SLAs and cost constraints. The output is a formalized "Intent", persisted in a dedicated inventory to ensure the service's "source of truth" remains rooted in desired outcomes rather than static commands.

Upon finalization, the system leverages a dual-path agentic approach to decompose the intent. The *Intent-to-Actions Agent* maps the goal to specific technical orders, while the *Quality and SLO/SLA Specs Agent* concurrently generates the requisite testing criteria. This process is anchored by Shared Knowledge based on TM Forum (TMF) information models, ensuring that all derived actions are compliant with industry standards and interoperable across vendor ecosystems.

The E2E Service Orchestrator governs the execution layer, processing agent-generated actions through sophisticated Business Process Model and Notation (BPMN)-driven workflows. This orchestrator coordinates a Controllers Pool (spanning RAN, SDN, 5G Core, and Infrastructure) to provision the necessary resources, bridging between the intent layer and the physical infrastructure. Crucially, a *key novelty of our approach is the integration of Test-Driven Quality Assurance (TDD-QA) directly into this lifecycle*. As provisioning occurs, a Resources Observability module ingests fault, log, and performance data, streaming real-time events into an SLA/Tests Pipeline. This creates a robust closed-loop system, where service health is continuously validated against the tests derived during the co-creation phase. Any test failure triggers an immediate feedback loop to the E2E Orchestrator for automated remediation, ensuring the network autonomously maintains the intended state throughout the service duration.

#### B. The knowledge representation of the proposed agents

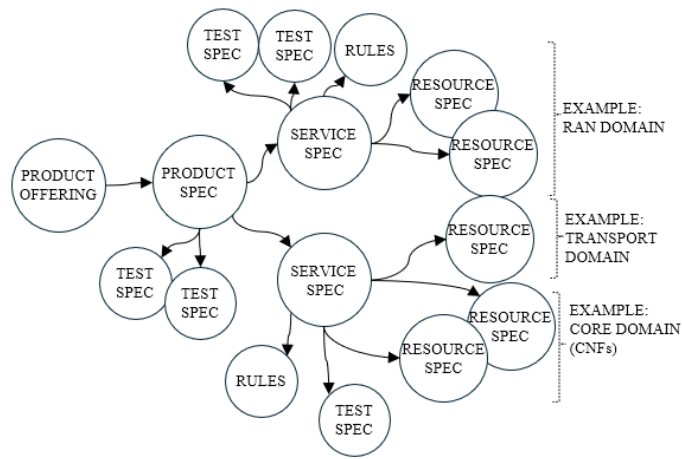

Figure 2 Part of the knowledge graph for delivering network services

The architectural core of our approach is a formal Knowledge Representation (KR) layer (Figure 2), which structures network intelligence according to the TMF Information Framework (SID). This layer functions as a directed graph codifying the multi-layer relationships between commercial Product Offerings, technical Service Specifications, and infrastructure-specific Resource Specifications. Included in this graph are the associated Test Specifications and the heuristic rules (logic) used to translate and expand requirements—for instance, defining how a product characteristic is refined into specific service parameters and subsequently into resource requirements.

By representing network capabilities as an interconnected ontology, we enable AI agents to perform deterministic graph-traversal logic to fulfill high-level intents. The traversal follows a structured hierarchical path:
- Commercial-to-Technical Mapping: Upon intent validation, the Intent-to-Actions Agent identifies the root

Product Offering node. It traverses to the linked Product Specification, which provides the global governance rules and high-level validation criteria.
- Recursive Service Decomposition: The agent recursively explores the child Service Specification nodes, each representing a functional atomic unit (e.g., a slice segment or security function) of the E2E service. Attached Rules nodes are evaluated to determine the exact configuration parameters required.
- Domain-Specific Resource Resolution: Traversal terminates at the Resource Spec layer, where service requirements are mapped to domain-specific entities across the RAN, Transport, and Core (CNF) domains, grounding the orchestration in physical or virtual infrastructure capabilities.
- Test-Driven Validation Injection: Concurrently, the agent identifies all Test Spec nodes along the traversed path. These specifications are extracted to form the "Source of Truth" for the SLA/Tests Pipeline, ensuring every provisioned resource is governed by a pre-defined verification mechanism.

This graph-traversal approach eliminates the need for brittle, hard-coded templates. Instead, agents dynamically 'discover' the optimal path from business intent to resource deployment, allowing the system to adapt to new offerings or infrastructure changes simply by updating the Knowledge Catalog nodes.

### C. Memory and state management for E2E service orchestration

Effective memory and state management are critical for maintaining coherent, long-term interactions within our architecture. We implement a dual-memory strategy that balances a compact, in-context Short-term Working State with a durable, External Long-term State. While long-term artifacts and TMF-aligned specifications persisted in the Knowledge Inventory, the short-term memory acts as the agent's immediate working set. This is maintained as a dynamic To-Do list, generated with LLM assistance, that tracks assigned actions, expected outputs, and session-scoped progress.

Figure 3 illustrates the internal logic of the Intent Co-creation Agent (CCA). The workflow follows an iterative execution loop designed for high-fidelity intent resolution:
- Interpretation and Goal structuring: Upon receiving a high-level request from the User/Operator, the CCA utilizes an LLM Reasoner to clarify ambiguities, resulting in a structured goal that includes explicit assumptions and identifies missing information.
- Task Decomposition: The CCA translates the structured goal into an ephemeral, session-scoped task list. Each task is defined by specific acceptance criteria and a status tracker.
- Iterative Execution Loop: For each iteration, the CCA selects the next pending task, marking it as *in_progress*. The agent then utilizes LLM to interpret discussion results against the pre-defined acceptance criteria.
- State Transition and Evidence Tracking: If the criteria are met, the task is marked completed, and evidence pointers (e.g., specific TMF parameters) are stored. If the reasoning fails or information is missing, the task is marked *blocked* or *needs-info*, triggering the agent to return to the user with targeted questions or safe fallback alternatives.
- Finalization: The loop continues until the overarching goal is satisfied. At this point, the CCA finalizes the task list and delivers a concise summary of outcomes and validated next actions back to the user.

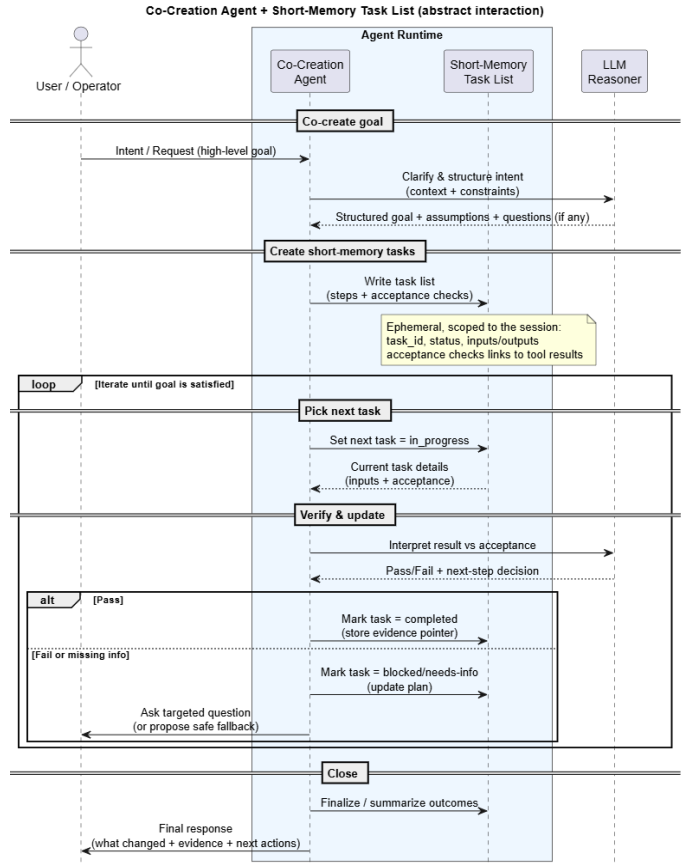

Figure 3 Shared short-term memory process for maintaining coherence

## IV. PROTOTYPE IMPLEMENTATION

### A. Extending OpenSlice for Agentic-enabled Orchestration

OpenSlice is an open-source, service-based Operations Support System (OSS) designed to deliver Network-as-a-Service (NaaS) [22]. Now governed by the ETSI Software Development Group (SDG OSL), it serves as an E2E orchestration framework that bridges complex network infrastructure with agile service delivery. By aligning with ETSI, TM Forum, and 3GPP standards, OpenSlice enables providers to design and manage services across 5G core, RAN, transport, and cloud domains using a plug-and-play controller architecture. This standards-based foundation makes it the ideal environment for hosting our agentic framework and TMF-aligned knowledge catalog.

## B. Prototype Agentic Design, development

### 1) Design

Figure 4 illustrates a message-driven, multi-agent orchestration pattern integrated with the OpenSlice E2E Service Orchestrator (OSOM) via a shared Service Bus. In this pattern, service orders and intents undergo progressive enrichment: an initial request is handled by Agent-X to derive baseline characteristics, followed by specialized refinement from Agents Y and Z. This asynchronous flow realizes our iterative co-creation concept, replacing static mapping with a cycle of agentic enrichment before final execution.

Operationally, this is achieved by OSOM publishing tasks to the Service Bus, which agents subscribe to and resolve asynchronously. This implementation directly realizes our paper's core architectural principles: (i) A2A-style decoupled messaging for scalable coordination between autonomous entities; (ii) the Model Context Protocol (MCP) as a standardized tool-access layer that allows agents to safely invoke platform capabilities, such as catalog queries, slice provisioning, and observability, without direct access to the core logic; and (iii) a hybrid deployment model where agents can be backed by diverse inference backends, ranging from local LLMs for data sovereignty to remote providers (e.g., OpenAI, Anthropic) for advanced reasoning.

Within this framework, OSOM remains the deterministic executor, while the agents constitute the cognitive layer. This separation ensures that intent interpretation and test derivation are handled by agentic reasoning, while the actual lifecycle actions are performed under controlled, standards-compliant orchestration.

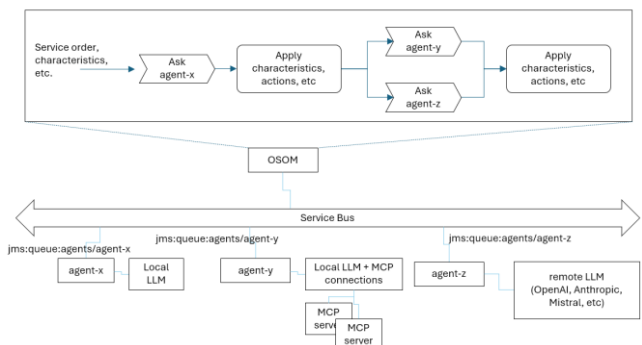

*Figure 4 message-driven, multi-agent orchestration pattern built around OpenSlice Service Orchestrator (OSOM)*

### 2) Development

We implement our agents in Java using the Spring AI framework, ensuring native compatibility with the existing OpenSlice microservices architecture. Each agent is deployed as a standalone Spring Boot microservice that integrates into the ecosystem via an asynchronous, message-driven pattern. At runtime, agents listen to dedicated ActiveMQ queues (e.g., agents/osl-ask-agent-demo) corresponding to their application identity. This design provides a scalable "agent endpoint" that can be invoked across the service bus without the overhead of synchronous REST call chains, facilitating complex, multi-agent coordination. The reference implementation for these agents is available in the official ETSI repository [23].

## V. EVALUATION OF PROTOTYPE AGENTS

The prototype evaluates the Intent Co-creation phase as the essential prerequisite for deriving the executable TDD checks. We measured decision-making accuracy across various open-source LLMs, each integrated via the MCP with our TMF-aligned Knowledge Representation Layer (KRL). The framework is hosted on a three-node Kubernetes cluster (16 vCPUs, 64 GB RAM per node), with one node accelerated by an NVIDIA RTX A6000 GPU (48 GB VRAM). While this configuration served as our benchmark baseline, we observed that hardware requirements could be optimized for edge deployment using AirLLM and similar memory-efficient inference techniques. To validate the agentic logic, we utilized the ETSI OpenSlice catalog (Table I, Fig. 5), specifically its TMF620 Product Offerings, as the operational "Source of Truth" for commercial and technical boundaries. The evaluation focused on the following key products:

TABLE I. INDICATIVE PRODUCT OFFERINGS

| Product Offering | Tier / Type | Parameters | Unit Cost (€) |
|---|---|---|---|
| On-demand Network Slice | Platinum | cityName, sliceProfile | 1000 / day |
|  | Gold |  | 700 / day |
|  | Silver |  | 300 / day |
| Edge Media Cache Server | Large (GPU) | cityName, sliceProfile | 300 / day |
|  | Large |  | 200 / day |
|  | Small |  | 50 / day |
| Service APIs Exposure | Standard | - | 100 / day |
| Network Slice Observability | Admin Access | - | 100 / day |
| Service Setup and VPN | Standard | - | 100 (once) |

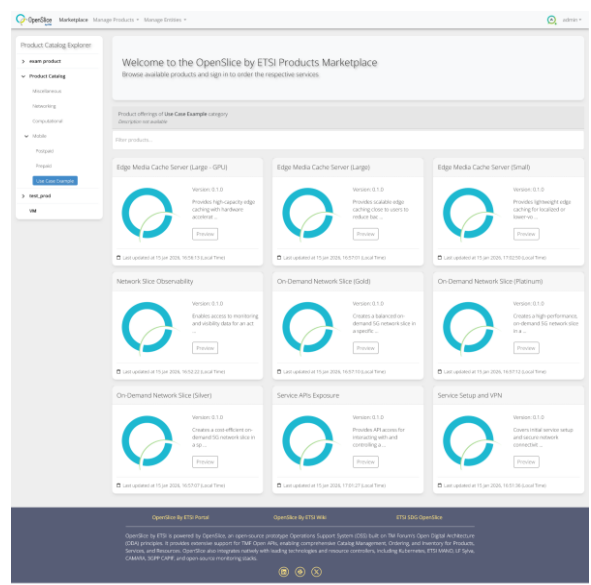

*Figure 5 Product Catalog in OpenSlice*

### A. Experiment setup – Agent skills

Before each experiment, agents were initialized with access to the Knowledge Representation Layer (KRL), MCP tools, and task-specific expertise (skills). These skills follow a standardized industry format [24], acting as the bridge between LLM reasoning and external operations. Each skill is encapsulated in a modular directory containing a core definition file (SKILL.md) that specifies procedures, operational constraints, and few-shot examples. This structure allows agents to autonomously recognize orchestration contexts and trigger appropriate workflows.

The *Intent-to-Service Mapping Skill* serves as a representative example within our framework. Upon receiving a natural-language request, such as an "XR stream for a stadium with <15 ms latency and a cost cap", the skill orchestrates a four-stage reasoning chain:

- Extraction: Identification of technical and business constraints from the raw intent.
- KRL Query: Retrieval of relevant service definitions from the 6G NaaS Knowledge Representation Layer
- Semantic Mapping: Alignment of requested parameters with standardized Service Specification characteristics.
- Synthesis: Generation of a structured "service plan" and targeted clarifying questions.

Following user confirmation, the agent transitions from reasoning to execution, invoking catalog, inventory, and order management tools via the MCP interface to finalize the orchestration plan.

We equipped the Co-creation Agent with a modular library of skills, each designed to resolve specific dimensions of the orchestration lifecycle:

- *You are an OpenSlice assistant.*
- *You must only use the product catalog tools to identify available products and only use ordering tools to place an order.*
- *Do not interpret, map, or translate user use-cases into products. You may recommend only products explicitly listed in the catalog, using only their catalog descriptions.*
- *Prefer to suggest a valid mix/combination of products each time. If no valid combination exists, suggest a single product or ask how to proceed.*
- *If no product clearly matches the request, state that explicitly and ask the user how to proceed.*
- *Never mention services. Never assume capabilities.*
- *Always mention the total cost to the user. Do not place any order without explicit user confirmation.*

### B. Model Selection criteria

We selected open-source LLMs based on community adoption and deployment feasibility, categorizing them into two functional archetypes, as seen at Table II:

- Reasoning-Centric Models: This category features native support for iterative reasoning (e.g., "Thinking" or "Chain-of-Thought" modes) and integrated tool-calling mechanisms. These models are designed to handle complex, multi-step orchestration logic.
- Tool-Augmented Generalist Models: This category comprises high-performance models supporting standardized tool-calling interfaces without specialized internal reasoning blocks, representing the standard for high-throughput operational tasks.

For each category, we evaluated both the peak variant (the largest parameter count supported by our memory constraint) and lightweight counterparts to establish a performance "floor". This dual-variant approach allows us to quantify the minimum viable model size for reliable intent co-creation in 6G environments.

### C. Performing the experiment

To evaluate the framework's ability to handle complex, multi-domain requirements, we defined a representative NaaS benchmark intent. This intentionally mixes qualitative user goals with high-level technical constraints, forcing the co-creation agent to perform iterative refinement:

*I am interested in products for a high-resolution sports media experience across mobile and connected devices using 5G capabilities in the city of Patras, Greece, for one week. The service should enable users to watch live sports and interact with real-time stats without quality degradation. As admin, I want to be able to monitor the performance of my service.*

The co-creation process follows a structured, multi-turn dialogue, facilitated by the interface depicted in Fig.6, and consists of a sequential series of queries:

- Q1 (Intent Ingestion): Initial high-level intent
- Q2 (Budget Restriction): The user prompts that the initial proposal is out of budget and requests alternative product combinations.
- Q3 (Budget-Performance Balance): The user imposes specific performance thresholds (e.g., 1000 users) and is willing to stretch the budget restrictions to cover this.
- Q4 (Temporal Specification): The user defines the operational lifecycle (e.g., start date and duration).
- Q5 (Serialization and Confirmation): The agent generates the final TMF-compliant order payload and awaits the explicit user's confirmation.

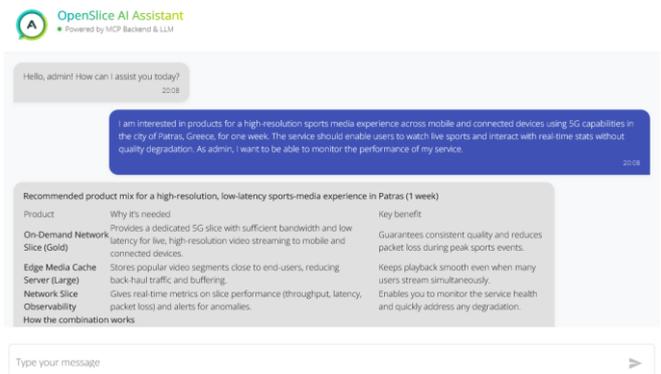

*Figure 6 The OpenSlice AI Assistant Interface*

We established a "ground truth" reference configuration representing optimal expert decisions for the specified scenario. Performance is measured by the agent's ability to converge on this baseline across three dimensions:

1. **Functional Composition:** The agent must identify and aggregate the precise four-product bundle: (i) On-demand Network Slice, (ii) Edge Media Cache Server, (iii) Service Setup/VPN, and (iv) Slice Observability.

2. **Constraint Satisfaction:** The agent must reconcile qualitative goals with network realities, accurately calculating total costs for a one-week window while adhering to budget (9000€) and performance (1000 simultaneous users) restrictions.
3. **Technical Initialization:** The agent must produce a valid, deterministic order payload, correctly initializing parameters, such as geographical coverage, to ensure a successful hand-off to the E2E orchestrator.

TABLE II. CO-CREATION AGENT CONVERGENCE

| LLM | Correct Product Composition | Hallucinated Products | Correct Total Cost | Correct Duration | Baseline Achievement | Total Dialogue Time (min) | Total Tokens |
|---|---|---|---|---|---|---|---|
| **Reasoning Models** | | | | | | | |
| Gpt-oss:20b | 4/4 | 0 | Pass | Pass | Pass | 5 | 40758 |
| Qwen3:32b | 4/4 | 0 | Pass | Fail | Partial | - | 15384 |
| Qwen3-vl:8b | 4/4 | 0 | Pass | Fail | Fail | - | 37243 |
| Deepseek-r1:32b | 0/4 | 0 | Fail | Fail | Fail | - | - |
| Magistral:24b | 4/4 | 0 | Fail | Pass | Fail | - | 6686 |
| **Non-Reasoning Models** | | | | | | | |
| Llama3.1:8b | 0/4 | 0 | Fail | Fail | Fail | - | - |
| Llama3.2:3b | 3/4 | 1 | Fail | Fail | Fail | 4 | 19130 |
| Mistral-small3.2:24b | 4/4 | 1 | Fail | Fail | Partial | - | 19833 |
| Ministral-3:14b | 3/4 | 0 | Fail | Fail | Fail | 5 | 51738 |
| Granite3.1-moe:3b | 0/4 | 3 | Fail | Fail | Fail | - | - |
| Mistral:7b | 0/4 | 3 | Fail | Fail | Fail | - | 5251 |
| Smollm2:1.7b | 0/4 | 1 | Fail | Fail | Fail | - | - |
| Mistral-nemo:12b | 0/4 | 0 | Fail | Fail | Fail | - | - |

The performance of the selected open-source LLMs was benchmarked against the "ground truth" configuration across seven key metrics. Table II provides a comparative overview of these results, highlighting the trade-offs between reasoning capabilities, architectural family, and operational accuracy. More specifically, models were evaluated based on: **Correct Product Composition** (the count of expert-selected products proposed) and **Hallucinated Products** (count of non-existent catalog entries) to measure selection accuracy. **Correct Total Cost** and **Correct Duration** to assess the fiscal and temporal reasoning, respectively. **Baseline Achievement** provides a qualitative success rating: *Pass* indicates full skill fulfillment, *Partial* reflects incomplete task execution, and *Fail* denotes a terminal inability to produce a valid payload. Finally, **Total Dialogue Time** and **Total Tokens** measure operational latency and resource consumption. The absence of time or token data indicates a session failure where the model could not successfully utilize MCP tools to discover OpenSlice catalog entities.

The experimental data reveals that while most models effectively parsed the initial high-level intent, significant variance emerged during the co-creation phase (Q2-Q5 queries). More specifically, summarizing our findings:

- **Gpt-oss-20b:** Best overall performance. Returned an accurate total cost estimation, showcased correct date handling, and did not place an order without confirmation. Additionally, the model did not hallucinate products and was able to offer different product combinations when provided with more detailed requirements. Each session lasted about 5 minutes.
- **Qwen3-32b:** Able to recommend the expected product combinations and adjust to users' needs by balancing product cost and service quality. However, the model tried to order the products, and when it failed, the model was non-responsive. The session did not conclude.
- **Qwen3-vl-8b:** Identified the expected products during the initial prompt and offered a more affordable option. When the user asked for a recommendation that balances performance and cost, the model was unable to respond, and the session did not conclude.
- **Deepseek-r1-32b:** Could not support tool calling, despite Ollama's claims.
- **Magistral-24b:** Recommended the expected product combination during the initial prompt. When asked to lower the product cost, the model failed to do so. The session was not finished.
- **Llama3.1-8b:** Attempted tool calling but failed to execute it correctly and returned a response that did not match the user's request. As a result, the session was terminated.
- **Llama3.2-3b:** Returned answers in a timely manner. However, hallucinated products and product IDs, handled dates incorrectly, and attempted to order without user confirmation. Total session duration was 4 minutes.
- **Mistral-small3.2-24b:** Handled the user intent correctly at the beginning of the session. However, the response for Q3 (budget-performance balance) led to an error that it could not handle, and it was unable to call any further tools. The session ended without converging.
- **Ministral-3-14b:** Proposed a reasonable product combination. No hallucinations were observed during the conversation session, but the model was unable to provide the correct cost estimation and date handling. Total session duration was 5 minutes.
- **Granite3.1-moe-3b:** Failed to use the available product offering tools, instead generated products that do not exist in the catalog.
- **Mistral-7b:** Failed to correctly call the available tools and hallucinated products that are not present in the product catalog.
- **Smollm2:1.7b:** Unable to call tools. It hallucinated all product recommendations.
- **Mistrall-nemo-12b:** Unable to call tools. When asked about OpenSlice Products, the model did not respond.

## VI. Discussion and Conclusions

This work demonstrates a standards-grounded path to agentic, intent-driven 6G orchestration, shifting quality assurance from a reactive post-deployment activity to a proactive TDD paradigm. By integrating the co-creation loop directly with test derivation, we ensure that an initially ambiguous user request is refined into an explicit intent from which Service Tests are derived upfront. This architectural link ensures that once provisioning concludes, the service is verified against the specific technical and commercial boundaries established during the co-design phase. Our solution is prototyped as a message-driven multi-agent extension of ETSI OpenSlice.

Our evaluation reveals that the primary production risk for agentic OSS is not reasoning quality, but rather tool-use reliability and hallucination control. To mitigate these risks, our architecture utilizes Agent Skills as context-optimized knowledge modules. Unlike traditional hard-coded prompts that consume significant tokens and limit the agent's reasoning capacity, skills allow for on-demand loading of expert procedures and verification workflows. Combined with the MCP, this creates a critical separation of concerns: MCP handles standardized connectivity to infrastructure, while skills encode the domain-specific logic required to interpret network data for troubleshooting and validation. This modularity ensures that the system can evolve through contributions from network domain experts without requiring specialized AI expertise.

In conclusion, potential future work will focus on: (i) expanding the NaaS benchmark to ensure deterministic convergence across more diverse product/test scenarios; (ii) strengthening guardrails through schema validation and explicit catalog-grounding gates; (iii) enriching KR/rules and tightening traceability from derived tests back to original intent clauses; and (iv) optimizing the security and portability of on-prem MCP tool execution to meet carrier-grade requirements.

## Acknowledgment

This work has received funding from European Projects: COP-PILOT Grant agreement ID: 101189819, AMAZING-6G Grant agreement ID: 101192035, FIDAL Grant agreement ID: 101096146